\documentclass{ifacconf}  

\usepackage{overpic}
\usepackage{amsfonts}
\usepackage{graphicx}
\usepackage{amsmath}
\usepackage{url}
\usepackage{ifthen}
\usepackage{color}
\usepackage{cite}
\usepackage{enumerate}
\usepackage{subfigure}
\usepackage{siunitx}
\usepackage{natbib}

\newtheorem{theorem}{Theorem}

\newtheorem{remark}{Remark}

\newcommand{\db}{d_{\mathrm{b}}}
\newcommand{\dt}{d_{\mathrm{t}}}
\newcommand{\diag}{\mathrm{diag}}
\DeclareSIUnit[]{\pu}{p.u.}

\begin{document}
\begin{frontmatter}

\title{Damping Ratio and Convergence Rate Analysis for Inverter-Based Power Systems via Root Locus\thanksref{footnoteinfo}}

\thanks[footnoteinfo]{This work was supported in part by National Natural Science Foundation of China under Grant 62573006 and in part by CUHKSZ University Development Fund 01003790.\\
(Corresponding author: Yan Jiang.)}

\author{Dan Wang$^{1}$, Wei Chen$^{2}$, Yan Jiang$^{3}$, and Shinji Hara$^{4}$}

\address{$^{1}$School of Robotics and Automation, Nanjing University, Suzhou, China (email: danwang@nju.edu.cn).}
\address{$^{2}$School of Advanced Manufacturing and Robotics, Peking
University, Beijing, China (e-mail: w.chen@pku.edu.cn).}
\address{$^{3}$School of Science and Engineering, The
Chinese University of Hong Kong, Shenzhen, China (e-mail:
yjiang@cuhk.edu.cn).}
\address{$^{4}$Institute of Integrated Research, Institute of Science Tokyo, Tokyo, Japan
(e-mail: shinjihara5202@gmail.com).}

\begin{abstract}
As more renewables enter the grid, it becomes insufficient to rely solely on the center of inertia frequency to analyze the transient performance, especially for oscillation behaviors. Motivated by this, we study the transient frequency performance of inverter-based power systems. Two key performance criteria are considered: the damping ratio and the convergence rate, which are determined by the angles and the real parts of the closed-loop poles, respectively. By decomposing the network system into a set of scalar feedback systems, we show that the closed-loop poles can be analyzed using the root locus of a scalar feedback system as the gain varies from the smallest nonzero eigenvalue to the largest eigenvalue of the scaled Laplacian matrix of the interconnection network. Based on this, we derive sufficient conditions that ensure desired damping and convergence performance. Our analysis shows that the worst-case damping ratio and convergence rate occur when the gain equals the largest and the smallest nonzero eigenvalues of the scaled Laplacian, respectively. This result significantly simplifies the performance assessment. We further provide guidelines for tuning the inverter parameter. Numerical simulations are included to illustrate and validate the proposed approach.

\end{abstract}


\end{frontmatter}

\section{Introduction}

Transient frequency performance following major disturbances is a
critical concern in power system operations. However, the fact that a power system usually has a variety of highly-coupled interaction among individual buses makes it nontrivial to analyze or tune the performance of post-contingency frequency responses~\citep{Khamisov2024TPS,JIANG2024epsr}. One approach to reduce the size of the problem is to study the center of inertia (COI) frequency, which is the inertia-weighted average of the frequency signals at individual buses~\citep{milano2017rotor,azizi2020local}. This approach works well when all buses exhibit a relatively coherent frequency response~\citep{Min2021lcss}. Leveraging inverter-based control, it is not hard to shape the COI frequency dynamics following a sudden power imbalance into a first-order one that naturally has no Nadir, which largely improves frequency security~\citep{jiangtps2021,jiang2021lcss}. However, as increased amount of low-inertia renewable resources located in remote areas enter the grid through long transmission lines, it becomes insufficient to only consider the COI frequency since pronounced inter-area oscillations can occur, which can cause minor inconvenience or catastrophic failures to system operation~\citep{JIANG2024epsr}. Thus, it is important to better understand how those oscillations are affected by the network properties and system parameters, which enables us to adjust inverter-based control to damp oscillations in a principled way.

In this paper, we investigate the transient frequency performance of inverter-based power systems, considering two key performance criteria: damping ratio and convergence rate. We adopt a proportional generator-inverter model~\citep{pm2019preprint,jiang2021tac}, which fits naturally into a homogeneous multi-agent  framework~\citep{Hara2014tac,HaraAuto2019}. Under this structure, the closed-loop poles are given by the poles of a family of scalar feedback systems that share the same loop transfer function but have different feedback gains, determined by the eigenvalues of a scaled Laplacian matrix. This enables us to use the root locus of a single scalar system, with the gain varying over the spectrum of the scaled Laplacian, to analyze the poles of all subsystems. Based on this root-locus characterization, we derive sufficient conditions for guaranteeing prescribed damping ratio and convergence rate. The conditions reveal a structural separation: the damping ratio depends only on the largest eigenvalue of the scaled network Laplacian, whereas the convergence rate is determined only by its smallest nonzero eigenvalue. This property significantly simplifies performance assessment. In addition, based on these conditions, we provide practical tuning guidelines for the controller parameter.



\section{Root locus method for multi-agent networks} \label{sec:problem}
In this section, we present our method within a general multi-agent framework that captures the essential feedback structure common to power systems and many other interconnected dynamical networks. The results developed here will serve as the foundation for the power-system case studied in the subsequent sections.

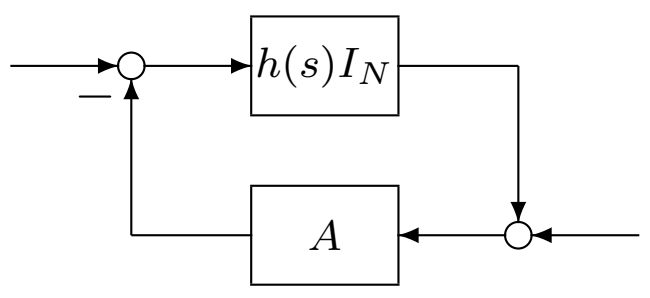
\begin{figure}[htb]
\begin{center}
\begin{picture}(53,22)
\put(0,18){\vector(1,0){9}}
\put(10,18){\circle{2}}
\put(11,18){\vector(1,0){9}}
\put(20,14){\framebox(12,8){\small $h(s)I_N$}}
\put(32,18){\line(1,0){10}}
\put(42,18){\vector(0,-1){13}}
\put(52,4){\vector(-1,0){9}}
\put(42,4){\circle{2}}
\put(41,4){\vector(-1,0){9}}
\put(20,0){\framebox(12,8){\small $A$}}
\put(20,4){\line(-1,0){10}}
\put(10,4){\vector(0,1){13}}
\put(7,14){\makebox(0,0)[b]{$-$}}
\end{picture}
\vspace{-5pt}
\caption{Feedback representation of a homogeneous multi-agent system.}
\label{fig:mas}
\end{center}
\end{figure}

Given a multi-agent system with $N$ homogeneous agents, where each agent is characterized by a transfer function $h(s)$ and the interconnection structure among the agents is specified by $A\in\mathbb{R}^{N\times N}$. 
This multi-agent system can be formulated as a feedback framework shown in Fig.~\ref{fig:mas}. Let
\begin{align*}
    h(s)I_N\#A:=\!\begin{bmatrix}
        (I_N+h(s)A)^{-1} & h(s)(I_N+h(s)A)^{-1} \\(I_N\!+\!h(s)A)^{-1}A & \ \ h(s)(I_N\!+\!h(s)A)^{-1}A  
    \end{bmatrix}\,.
\end{align*}
The multi-agent system is stable if the poles of $h(s)I_N\#A$ are contained in the open left half complex plane.

\begin{figure}[t]
\centering
\includegraphics[width=0.32\columnwidth]{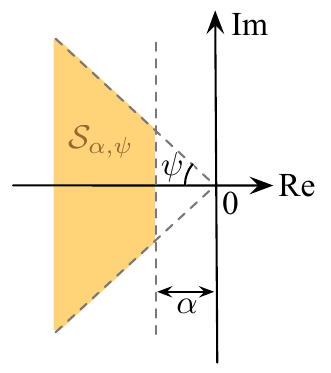}
\vspace{-10pt}
\caption{Region $\mathcal{S}_{\alpha,\psi}$.}
\label{fig:S-region}
\end{figure}

Such feedback framework provides a unified foundation for stability and performance analysis of a multi-agent system so that the methods from feedback control theory can be exploited. 
Two key performance criteria are of particular interest in this paper: the damping ratio and the convergence rate, which are determined by the angles and the real parts of the closed-loop poles, respectively.
As illustrated in Fig.~\ref{fig:S-region}, if all poles of $h(s)I_N\#A$ lie in the region $\mathcal{S}_{\alpha,\psi}$ with $\alpha>0$ and $\psi\in[0,\pi/2)$, defined as the intersection of a left-pointing sector of opening angle $2\psi$ and a vertical half plane with distance $\alpha$ to the left of the imaginary axis, then the outputs of the multi-agent system decay exponentially to zero with damping ratio no less than $\cos\psi$ and convergence rate at least $\alpha$. 
It is of importance to derive conditions under which the multi-agent network is stable and satisfies these prescribed convergence rate and damping ratio requirements.

According to \citep[Lemma 1]{Hara2014tac}, the closed-loop poles of $h(s)I_N\#A$ are given by the union of the closed-loop poles of $N$ independent scalar subsystems $$h(s)\# \lambda_i(A)\,,\qquad i=1,\dots, N\,,$$ where $\lambda_i(A)$ denote the eigenvalues of $A$. Hence, the performance of the multi-agent network in Fig.~\ref{fig:mas} can be analyzed by checking the corresponding performance of above scalar subsystems.
If 
$
h(s)=n(s)/d(s)
$,
where $n(s)$ and $d(s)$ are coprime, then
the characteristic polynomial of the subsystem $h(s)\#\lambda_i(A)$ is $\Phi_i(s):=d(s)+\lambda_i(A) n(s)$.

In this paper, we assume that $A$ is symmetric, yielding that all of its eigenvalues are real. Then the poles of $h(s)\# \lambda_i(A)$ can be analyzed by using the root locus of 
$\Phi(s):=d(s)+kn(s)$
when $k$ varies over  $[\underline\lambda(A), \overline\lambda(A)]$, where $\overline\lambda(A)$ and $\underline\lambda(A)$ denote the largest and smallest eigenvalues of $A$, respectively. In applying the root locus method, the key lies in identifying when the damping ratio and the convergence rate reach their worst-case values, respectively, as $k$ varies over this interval, which is generally nontrivial. To our delight, for the power system to be considered in the next section, such worst-case can be determined analytically due to the special structure of feedback control design.

\begin{remark}
For the convergence rate, all roots of $\Phi(s)$ should lie in $\mathcal{S}_{\alpha, \pi/2}$, which is obtained by shifting the left half-plane leftward by $\alpha$. Thus, the desired convergence rate can be checked by applying the standard continuous-time stability criterion to $\Phi(s-\alpha)$. Equivalently, we can shift the root locus of $\Phi(s)$ rightward by $\alpha$ and check the stability of the resulting system. 
Similarly, for the damping ratio, all roots of $\Phi(s)$ should lie in {$\mathcal{S}_{0,\psi}\cap \mathcal{S}_{0,-\psi}$},
which are derived by rotating the left half complex plane counterclockwise and clockwise by $(\pi/2-\psi)$ around the origin, respectively. Since $\mathcal{S}_{0,\psi}$ and $\mathcal{S}_{0,-\psi}$ are symmetric with respect to the real axis, it suffices to consider only one of them. Thus, the desired damping ratio can be checked by applying the standard stability criterion to $\Phi(e^{j(\pi/2-\psi)}s)$. Equivalently, we can rotate the root locus of $\Phi(s)$ counterclockwise by $(\pi/2-\psi)$ around the origin and check the stability of the resulting system. 
\end{remark}

\section{Modeling of power systems with motivations}
\label{sec:model}
In this section, we introduce the inverter-based power-system model that fits the multi-agent framework developed in Section \ref{sec:problem} and the motivations for studying the transient performance of power systems.

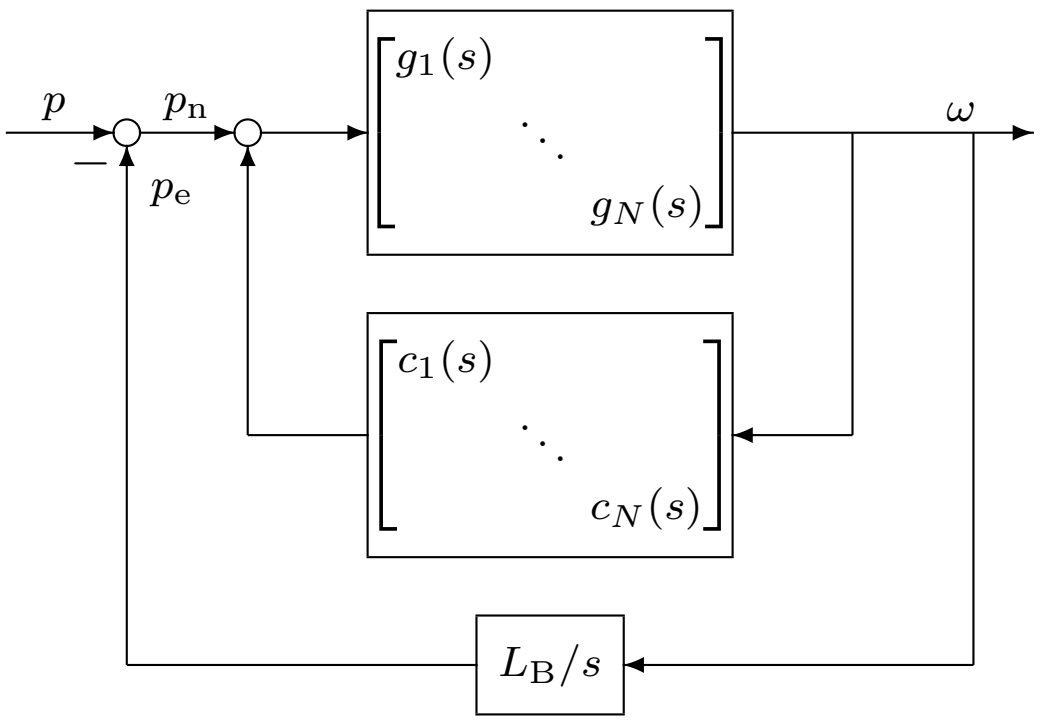
\begin{figure}[htb]
\begin{center}
\begin{picture}(85,58)
\put(0,48){\vector(1,0){9}}
\put(4,49){\makebox(0,0)[b]{\small$p$}}
\put(10,48){\circle{2}}
\put(11,48){\vector(1,0){8}}
\put(15,49){\makebox(0,0)[b]{\small$p_\mathrm{n}$}}
\put(20,48){\circle{2}}
\put(21,48){\vector(1,0){9}}
\put(30,38){\framebox(30,20){\small {$\begin{bmatrix}
g_1(s)&&\\&\ddots&\\&&g_N(s)
\end{bmatrix}$}}}
\put(60,48){\vector(1,0){25}}
\put(79,49){\makebox(0,0)[b]{\small$\omega$}}
\put(70,48){\line(0,-1){25}}
\put(70,23){\vector(-1,0){10}}
\put(30,13){\framebox(30,20){\small {$\begin{bmatrix}
c_1(s)&&\\&\ddots&\\&&c_N(s)
\end{bmatrix}$}}}
\put(30,23){\line(-1,0){10}}
\put(20,23){\vector(0,1){24}}
\put(80,48){\line(0,-1){44}}
\put(80,4){\vector(-1,0){29}}
\put(39,0){\framebox(12,8){\small ${L_{\mathrm{B}}/s}$}}
\put(39,4){\line(-1,0){29}}
\put(10,4){\vector(0,1){43}}
\put(12,43){\makebox(0,0)[l]{\small$p_\mathrm{e}$}}
\put(7,44){\makebox(0,0)[b]{$-$}}
\end{picture}
\vspace{-5pt}
\caption{Linearized power system model.}
\label{fig:power diagram}
\end{center}
\end{figure}

A power network can be modeled by an undirected connected graph $\mathbb{G}=(\mathcal{V},\mathcal{E})$, where $\mathcal{V}:=\{1,\dots, N\}$ represents the set of buses and $\mathcal{E}$ represents the set of transmission lines. The linearized model around an equilibrium can be represented as a feedback interconnection of bus dynamics and network dynamics, as illustrated in Fig.~\ref{fig:power diagram}. The input and output are the
power disturbances $p\in\mathbb{R}^N$ and the per unit bus frequency deviations from the nominal value $\omega\in\mathbb{R}^N$, respectively.

The dynamics of bus $i$ consists of the feedback interconnection of a generator $g_i(s)$ and an inverter $c_i(s)$. 
The network dynamics is given by a linearized model of the power flow equations \citep{Purchala2005dc-flow}: $    p_\mathrm{e}(s)=\frac{L_{\mathrm{B}}}{s}\omega(s),$
where $p_\mathrm{e}\in\mathbb{R}^n$ represents the fluctuations in power drained into the transmission network and $L_{\mathrm{B}}\in\mathbb{R}^{N\times N}$ is an undirected Laplacian matrix of the network whose $ij$th entry is:
\[
L_{\mathrm{B},{ij}}:=\Omega_0\partial_{\theta_j}{\sum_{l=1}^n|V_i||V_l|B_{il}\sin(\theta_i-\theta_l)}\Bigr|_{\theta={\theta}_0}\,.
\]
Here, $\theta=\begin{bmatrix}
    \theta_1 & \cdots &\theta_N
\end{bmatrix}^T\in \mathbb{R}^N$ is the voltage angle of the buses with $\theta_0\in\mathbb{R}^N$ being the equilibrium angles, $|V_i|$ is the (constant) voltage magnitude at bus $i$, $B_{ij}$ is the susceptance of line $(i,j)$, and $\Omega_0$ is the nominal frequency.

In this paper, we adopt the proportional model of the generators and the inverters  \citep{pm2019preprint}: 
\begin{align}
  g_i(s)=\frac{g_0(s)}{r_i} \qquad\text{and}\qquad
  c_i(s)=r_i c_0(s)\,,\label{eq:propc}
\end{align}
where $r_i>0$ for all $i\in\mathcal{V}$ are proportional parameters, and $g_0(s)$ and $c_0(s)$ are called the representative generator and the representative inverter, respectively.

It is well-established that under proportionality assumption, the common behavior of all buses is determined by~\citep{pm2019preprint,jiang2021tac,JIANG2024epsr} 
\begin{align}\label{eq:q(s)}
q(s):=\frac{g_0(s)}{1-g_0(s)c_0(s)}\,,   
\end{align} 
which is essential in power system analysis, as it characterizes the COI frequency response. The COI frequency, defined as the weighted average of all bus frequencies, is often regarded as a reliable representative of the overall system frequency when the power network is tightly connected \citep{Min2021lcss}. 

A typical representative generator is of the form
\begin{equation}\label{eq:go}
g_0(s)=\frac{ \tau s + 1 }{m \tau s^2 + \left(m + d \tau \right) s + d + \dt},   
\end{equation}
where $m>0$ is the representative generator inertia, $d>0$ is the representative generator damping, $\tau>0$ the representative turbine time constant, and $\dt>0$ is the representative turbine inverse droop.

Our prior work \citep{jiang2021tac,jiangtps2021,jiang2021lcss} has proposed frequency shaping (FS) control which can entirely remove the COI frequency Nadir to greatly improve the frequency security. Thus, we employ FS control for inverters here as a starting point for transient performance analysis, i.e.,
\begin{equation} \label{eq:dy-fs}
c_0(s) = \frac{ d_\mathrm{t} }{\tau s+1}-\left(d_\mathrm{b} + d_\mathrm{t}\right) \,,
\end{equation}
where $d_\mathrm{b}>0$ is a tunable parameter. 

A key advantage of FS controller in \eqref{eq:dy-fs} is that, for a system with $g_0(s)$ in \eqref{eq:go}, it renders $q(s)$ in \eqref{eq:q(s)} into first-order, i.e,
\begin{align}\label{eq:q(s)-fs}
q(s)=\frac{1}{ms + d + \dt+ \db}\,,   
\end{align}
which naturally ensures that the COI frequency exhibits no oscillations following a sudden step disturbance $p$.

However, relying solely on the COI frequency is insufficient. As renewable resources are increasingly integrated, especially at edges of the power network, pronounced oscillatory behavior may arise \citep{Khamisov2024TPS}. In such cases, the frequency
trajectories of individual buses may exhibit substantial inter-area oscillations following disturbances, even when the COI frequency remains well behaved \citep{JIANG2024epsr}. This may cause equipment damage or even catastrophic failures. Therefore, it is essential to characterize and ensure adequate damping and convergence for all components of $\omega$.
This motivates the present study, which aims to guarantee prescribed damping ratio and convergence rate for all trajectories of $\omega$. 

\begin{figure}[htb]
\begin{center}
\begin{picture}(86,22)
\put(0,18){\vector(1,0){9}}
\put(4,19){\makebox(0,0)[b]{\small$p$}}
\put(10,18){\circle{2}}
\put(11,18){\vector(1,0){8}}
\put(19,14){\framebox(12,8){\small $R^{-\frac{1}{2}}$}}
\put(31,18){\vector(1,0){6}}
\put(37,14){\framebox(12,8){\small {$q(s)I_N$}}}
\put(49,18){\vector(1,0){6}}
\put(55,14){\framebox(12,8){\small $R^{-\frac{1}{2}}$}}
\put(67,18){\vector(1,0){18}}
\put(79,20){\makebox(0,0)[b]{\small$\omega$}}
\put(76,18){\line(0,-1){14}}
\put(76,4){\vector(-1,0){27}}
\put(37,0){\framebox(12,8){\small ${L_{\mathrm{B}}/s}$}}
\put(37,4){\line(-1,0){27}}
\put(10,4){\vector(0,1){13}}
\put(7,14){\makebox(0,0)[b]{\small$-$}}
\end{picture}
\vspace{-5pt}
\caption{The proportional model.}
\label{fig:proportional}
\end{center}
\end{figure}
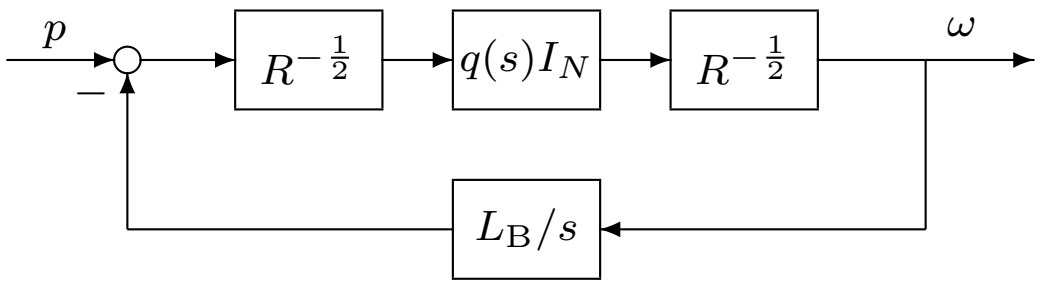

Under the proportional model \eqref{eq:propc} with the representative generator \eqref{eq:go} and the FS controller \eqref{eq:dy-fs}, the feedback system in Fig.~\ref{fig:power diagram} can be transformed into the feedback interconnection shown in Fig.~\ref{fig:proportional}, where $R:=\diag\{r_1,\dots, r_N\}$.
Define the scaled Laplacian matrix by  
$
L:=R^{-\frac{1}{2}}L_{\mathrm{B}}R^{-\frac{1}{2}}
$, which has a simple zero eigenvalue and all other eigenvalues are positive, i.e., $0 = \lambda_1 (L)< \lambda_2 (L)\leq \cdots \leq\lambda_N(L)$~\citep{pm2019preprint,jiang2021tac,JIANG2024epsr}. Then, using similar arguments in the proof of \citep[Lemma 1]{Hara2014tac}, we can show that the closed-loop poles of the system shown in Fig.~\ref{fig:proportional} are given by the union of the poles of the scalar subsystems 
\begin{align}\label{eq:subsystems}
 q(s)\#\frac{\lambda_i(L)}{s}\,,\qquad\ i\in\mathcal{V}\,.   
\end{align}
For $\lambda_1(L)=0$, the feedback branch in Fig.~\ref{fig:proportional} is opened,
and the transfer function from $p$ to $\omega$ reduces to $q(s)$ in \eqref{eq:q(s)}. 
Since $q(s)$ is first-order that naturally has no oscillations, the damping ratio of the subsystem associated with the zero eigenvalue requires no further examination. Moreover, as will be shown in the subsequent section, this subsystem achieves a convergence rate no smaller than that of any subsystem corresponding to a nonzero eigenvalue. 
Therefore, in the remainder of our analysis we focus exclusively on the subsystems associated with the nonzero eigenvalues of $L$.

\section{Transient performance analysis for power systems by root locus method}
\label{sec:power}

\subsection{Conditions for achieving required performance}

\begin{figure}[htb]
    \centering
    \subfigure[Root locus]{
    \includegraphics[scale=0.35]{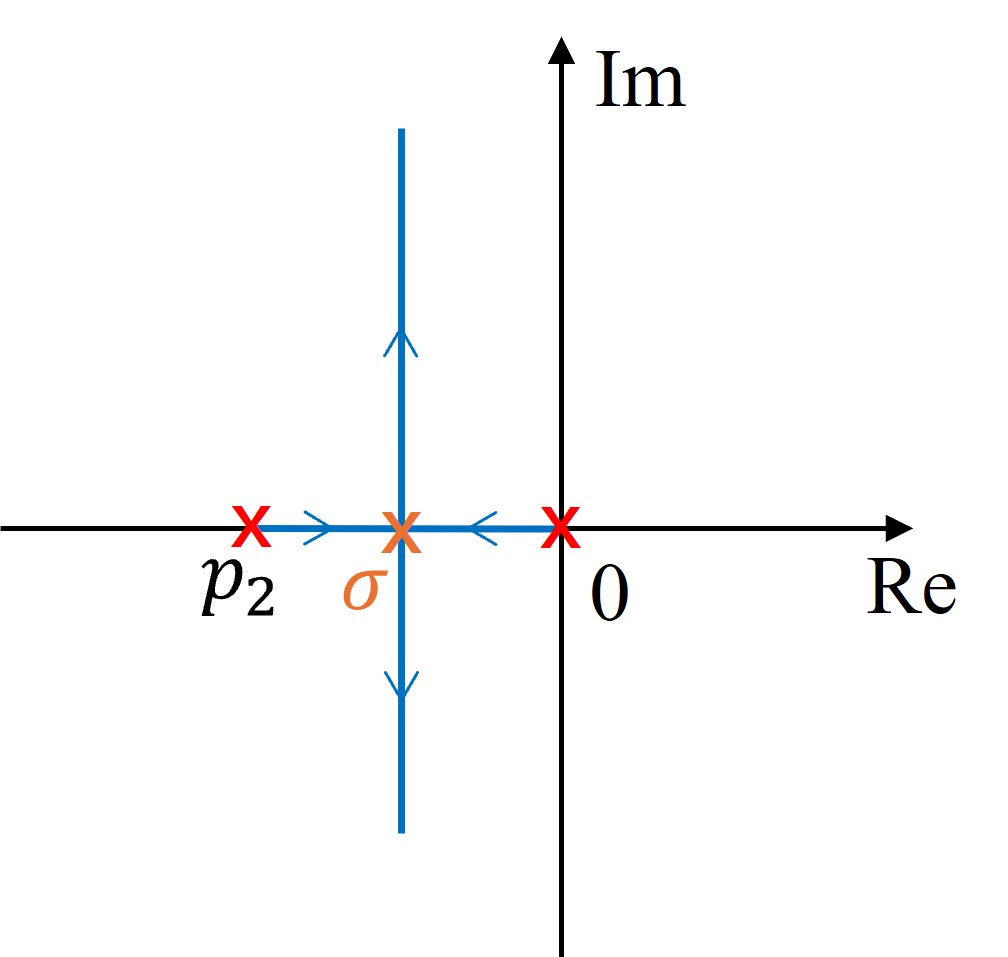}
    }
    \subfigure[Damping ratio]{
    \includegraphics[scale=0.35]{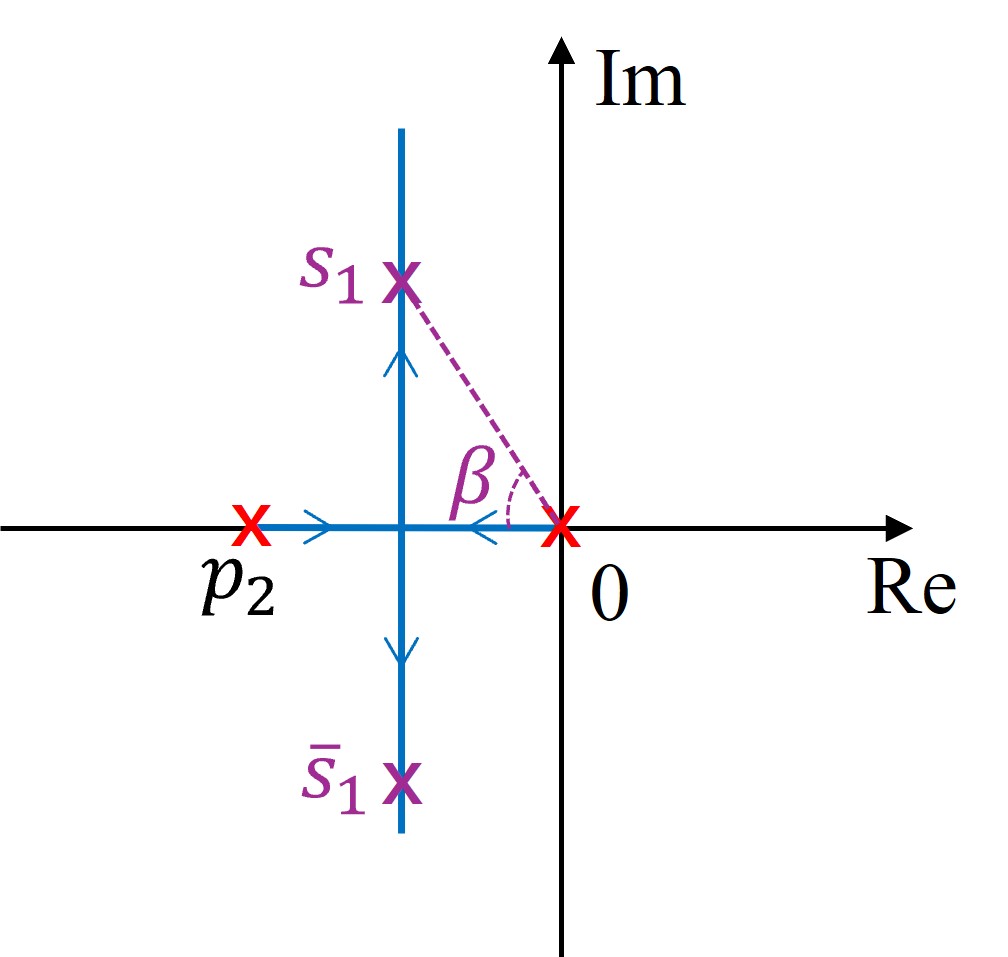}}
        \subfigure[Convergence rate]{
    \includegraphics[scale=0.35]{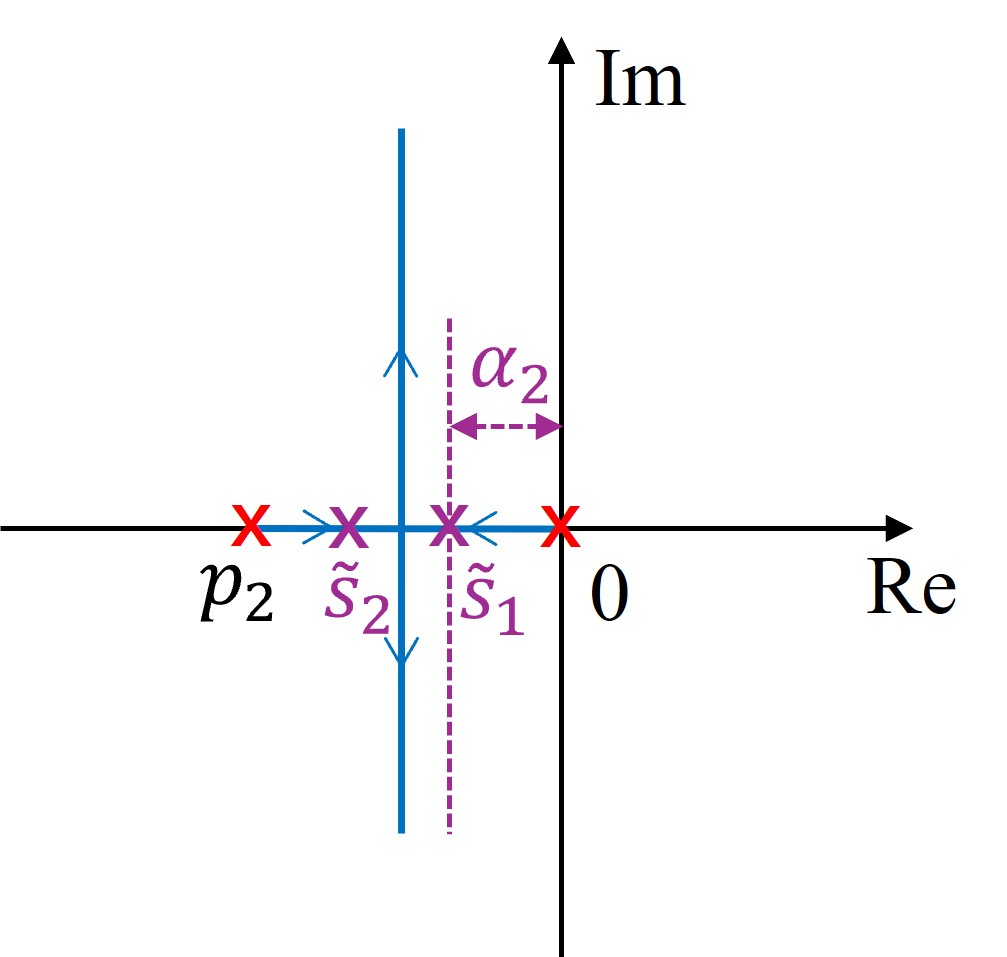}}
    \vspace{-10pt}
    \caption{Transient performance analysis by root locus.}
    \label{fig:rlocusdamping}
\end{figure}

In this subsection, 
we analyze the closed-loop poles of the subsystems in \eqref{eq:subsystems} that correspond to nonzero eigenvalues of $L$. Our goal is to derive conditions on the controller parameter $\db$ for FS control in \eqref{eq:dy-fs}, under which the power system in Fig.~\ref{fig:proportional} satisfies the prescribed damping ratio and convergence rate.
To this end, we group $1/s$ with $q(s)$ in \eqref{eq:q(s)} and let 
\begin{align}\label{eq:h(s)}
h(s):=\dfrac{q(s)}{s}=\dfrac{1}{ms^2+(d+\dt+\db)s}\,.
\end{align}
Then the closed-loop poles of \eqref{eq:subsystems} coincide with those of $h(s)\# \lambda_i(L)$ for all $i\in\mathcal{V}\setminus\{1\}$.
We can examine the poles of these subsystems $h(s)\# \lambda_i(L)$ by studying the root locus of the system $h(s)\#k$ when $k$ varies over  $[\underline\lambda(L), \overline\lambda(L)]$, whose associated characteristic polynomial is
\begin{align}\label{eq:characteristic}
    \Phi(s)=ms^2+(d+\dt+\db)s+k\,.
\end{align}
Here, with abuse of notation, we define $\underline\lambda(L):=\lambda_2(L)$ and $\overline\lambda(L):=\lambda_N(L)$ in the remainder of this paper.
As shown in Fig.~\ref{fig:rlocusdamping}(a), root locus of $\Phi(s)$ in \eqref{eq:characteristic} starts from two poles of $h(s)$: $p_1=0$, $p_2=-(d+\dt+\db)/m.
$
The breakaway point, coinciding with the intersection point of the asymptotes with real axis, is given by 
\begin{align}\label{eq:sigma}
\sigma=\frac{p_1+p_2}{2}=-\frac{d+\dt+\db}{2m}\,,    
\end{align}
with corresponding gain  
\begin{align}\label{eq:ksig}
k_{\sigma}=\dfrac{(d+\dt+\db)^2}{4m}\,.    
\end{align}




1) {\em Damping ratio analysis}: 

It is straightforward to see that if all closed-loop poles 
lie on the negative real axis, or equivalently, if $\overline\lambda(L)\leq k_{\sigma}$, then there will be no oscillation in the trajectories of $\omega$. 

If $\overline\lambda(L)> k_{\sigma}$, the damping ratio is determined by the closed-loop poles when $k=\overline{\lambda}(L)$ as illustrated in Fig.~\ref{fig:rlocusdamping}(b). These poles can be computed by setting $\Phi(s)=0$ with $k=\overline{\lambda}(L)$ in \eqref{eq:characteristic} as
\[
{s}_1,\bar{{s}}_1=\frac{-(d+\dt+\db)\pm j\sqrt{4m\overline\lambda(L)-(d+\dt+\db)^2}}{2m}\,.
\]
Using geometric relations, the damping ratio is given by
\begin{align}\label{eq:dampingexpression}
\cos\beta=\frac{|\sigma|}{|{s}_1|}=\frac{d+\dt+\db}{\sqrt{4m\overline\lambda(L)}}\,.    
\end{align}
Therefore, for a given $\psi\in [0,\pi/2)$, if 
\begin{align}\label{eq:dampingcon}
    \db\geq \cos\psi \sqrt{4m\overline\lambda(L)} -d-\dt\,,
\end{align}
then the damping ratio of the system is at least $\cos\psi$.

2) {\em Convergence rate analysis}: 

Note that $q(s)$ in \eqref{eq:q(s)-fs} has only one pole at $p_2$, which lies at the leftmost position on the root locus. Consequently, the subsystem in \eqref{eq:subsystems} corresponding to the zero eigenvalue of $L$ achieves the largest convergence rate, i.e., $|p_2|$, among all subsystems, as  mentioned at the end of Section \ref{sec:model}.

Let us analyze the convergence rate of subsystems in \eqref{eq:subsystems} corresponding to nonzero eigenvalues, where $\underline\lambda(L)$ plays a critical role. There are two possible cases:
\begin{enumerate}[i)]
    \item If $\underline\lambda(L)\geq k_{\sigma}$, then all closed-loop poles lie on the asymptotes. In this case, the convergence rate is 
    \begin{align}\label{eq:alpha1}
     \alpha_1=-\sigma= \frac{d+\dt+\db}{2m}\,.  
    \end{align}
    \item If $\underline\lambda(L)< k_{\sigma}$, then the convergence rate is determined by the poles when $k=\underline\lambda(L)$, which are 
    \[
    \tilde{s}_1,\tilde{s}_2\!=\!\frac{-(d\!+\!\dt\!+\!\db)\!\pm\!\sqrt{(d+\dt+\db)^2-4m\underline\lambda(L)}}{2m}\,,
    \]
    as shown in Fig.~\ref{fig:rlocusdamping}(c). Thus, the convergence rate is 
    \begin{align}\label{eq:alpha2}
        \alpha_2\!=&-\tilde{s}_1\nonumber\\=&\ \frac{d\!+\!\dt\!+\!\db\!-\!\sqrt{(d\!+\dt\!+\!\db)^2\!-\!4m\underline\lambda(L)}}{2m}\,.
    \end{align}
\end{enumerate}

By combining the two cases above, we conclude that the frequency $\omega$ decays exponentially to steady state with convergence rate at least $\alpha>0$ if either 
\begin{align}\label{eq:ratecon1}
    \underline\lambda(L)\geq k_{\sigma}\,, \ -\sigma\geq \alpha\,,
\end{align}
or
\begin{align}\label{eq:ratecon2}
\underline\lambda(L)< k_{\sigma}\,,\  \alpha_2 \geq \alpha\,.
\end{align}
holds. Simple calculations yield that \eqref{eq:ratecon1} is equivalent to
\begin{align}\label{eq:ratecon11}
2m\alpha -d-\dt<\db\leq \sqrt{4m\underline\lambda(L)}-d-\dt    
\end{align}
and \eqref{eq:ratecon2} is equivalent to
\begin{align}\label{eq:ratecon22}
\sqrt{4m\underline\lambda(L)}-d-\dt \leq \db \leq \frac{m\alpha^2+\underline\lambda(L)}{\alpha}-d-\dt\,.  
\end{align}

Summarizing the analysis for damping ratio and convergence rate, we obtain the following result: 
\begin{theorem}\label{thm:main}
Let $\psi\in[0,\pi/2)$ and $\alpha>0$. The frequency $\omega$ under FS decays exponentially to steady state with convergence rate at least $\alpha$ and damping ratio no less than $\cos\psi$ if either \eqref{eq:dampingcon} and \eqref{eq:ratecon11} or \eqref{eq:dampingcon} and \eqref{eq:ratecon22} are satisfied.  
\end{theorem}

Theorem~\ref{thm:main} reveals a useful structural property of the system under FS: the damping ratio depends only on the largest eigenvalue of $L$. In contrast, the convergence rate is determined only by the smallest nonzero eigenvalue of $L$. Thus, the two performance metrics are governed by different extremal components of the network spectrum. This observation significantly simplifies performance assessment of the power sytem under FS. That is, the damping ratio and convergence rate attain their worst-case values when $k=\overline{\lambda}(L)$ and $k=\underline{\lambda}(L)$, respectively. Consequently, evaluating transient performance reduces to analyzing these two scalar subsystems.

\subsection{Guidelines for tuning controller parameter}

Theorem~\ref{thm:main} relates the controller parameter $\db$ to the required convergence rate and damping ratio specifications. Based on this result, we next examine how $\db$ affects the transient performance and provide practical guidelines for its selection under given performance specifications.

\begin{figure}[htb]
    \centering
    \includegraphics[scale=0.08]{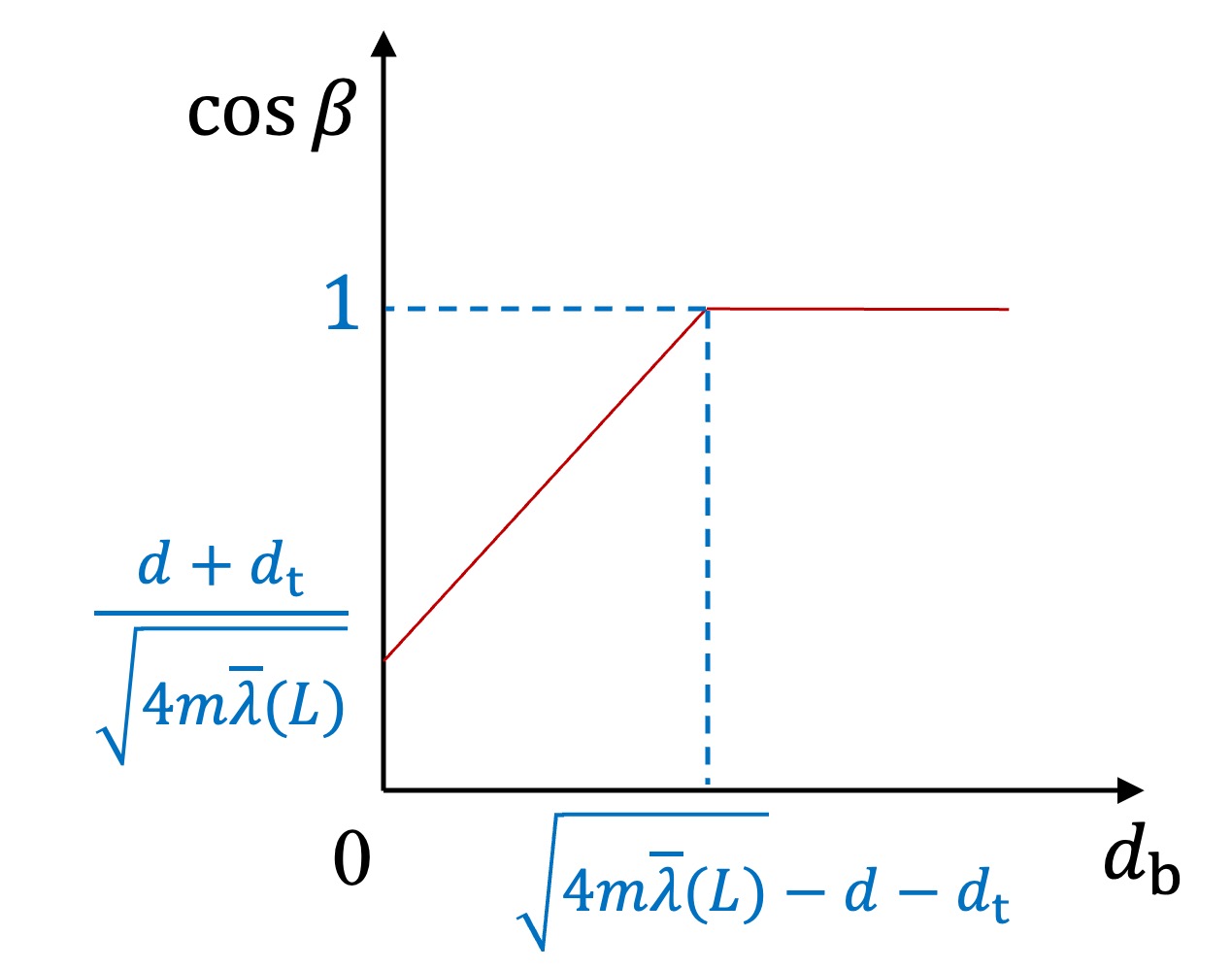}
    \caption{Relationship between $\cos\beta$ and $\db$.}
    \label{fig:db_vs_damping}
\end{figure}

We first explore how the damping ratio $\cos\beta$ varies with the controller parameter $\db$.
From \eqref{eq:dampingexpression}, $\cos\beta$ increases monotonically as $\db$ increases from zero, which is consistent with our intuition since increasing $\db$ adds damping to the system $h(s)$ in \eqref{eq:h(s)}. Meanwhile, the breakaway gain $k_{\sigma}$ in \eqref{eq:ksig} also increases with $\db$. As depicted in Fig.~\ref{fig:db_vs_damping}, when $\db$ is sufficiently large such that $k_{\sigma}=\overline\lambda(L)$, the damping ratio reaches $\cos\beta=1$; further increasing $\db$ does not improve it. This critical value of $\db$ is
$$ \sqrt{4m\overline\lambda(L)}-d-\dt\,.$$

\begin{figure}[htb]
    \centering
    \subfigure[$\underline\lambda(L)< \hat{k}_{\sigma}$]{
    \includegraphics[scale=0.4]{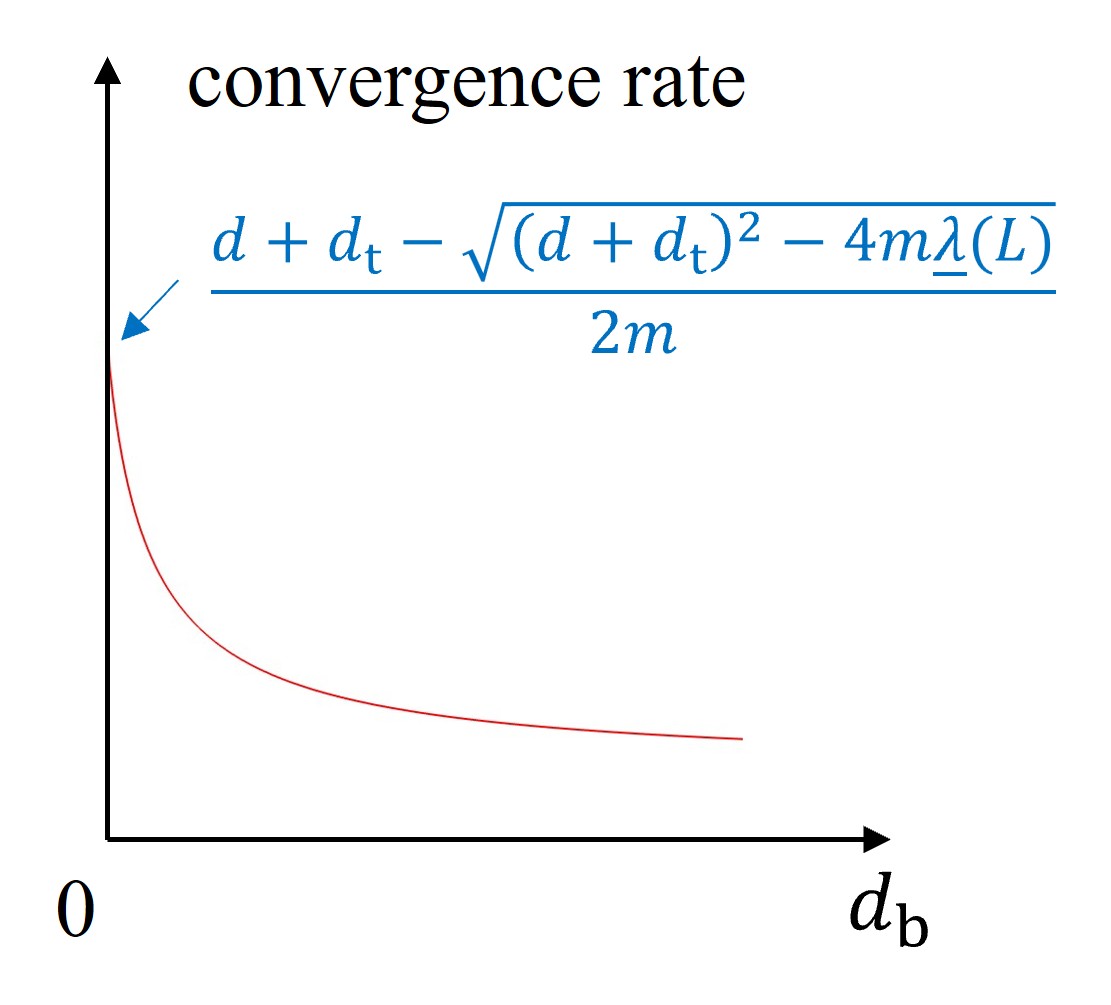}}
    \subfigure[$\underline\lambda(L)\geq \hat{k}_{\sigma}$]{
    \includegraphics[scale=0.4]{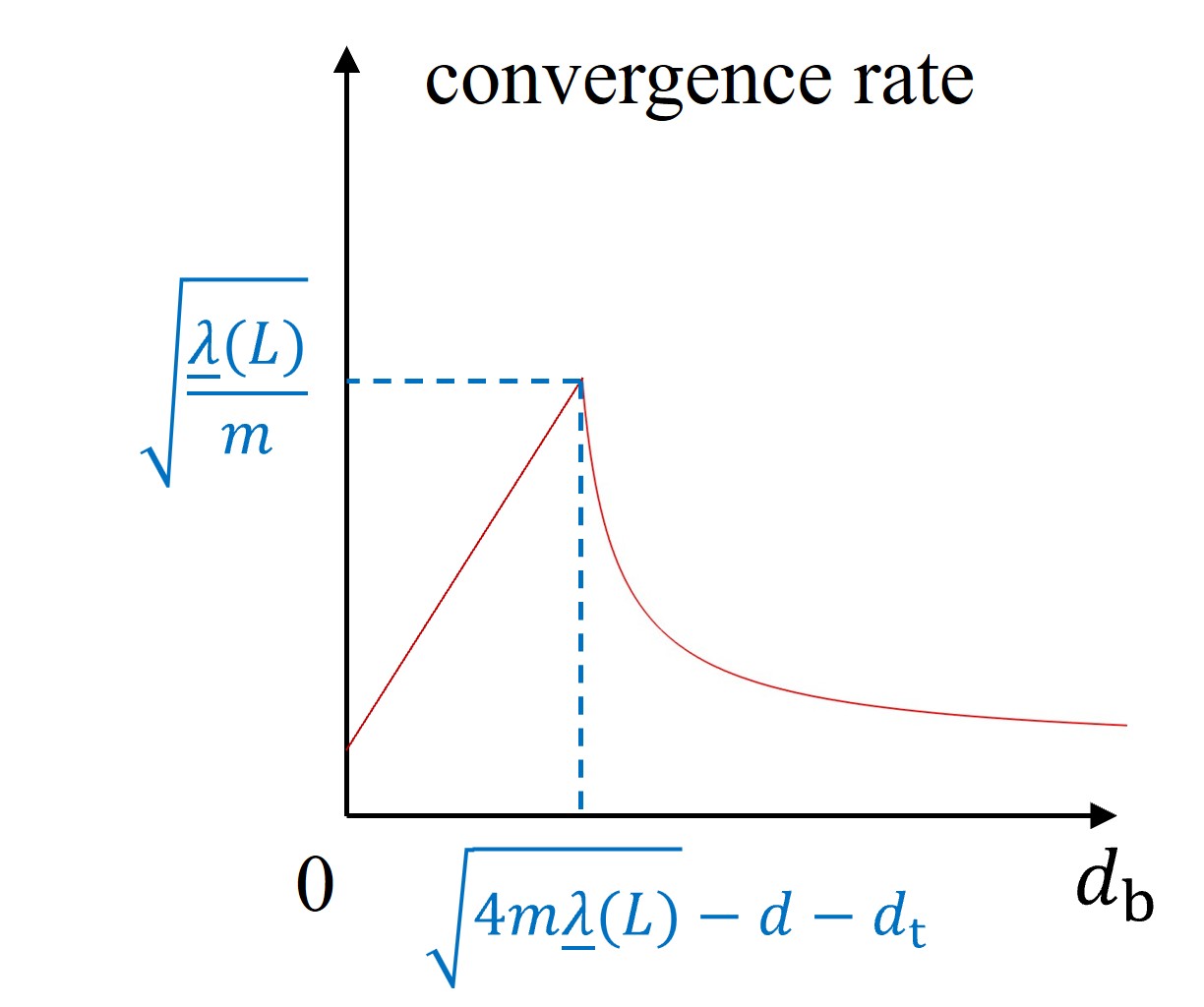}}
    \caption{Relationship between convergence rate and $\db$.}
    \label{fig:db_vs_rate}
\end{figure}

Next, we analyze how the convergence rate varies with $\db$.
If $\db=0$, the breakaway point of the root locus in \eqref{eq:sigma} is
$\hat\sigma=-(d+\dt)/(2m)$ and the corresponding breakaway gain in \eqref{eq:ksig} becomes $\hat{k}_{\sigma}=(d+\dt)^2/(4m)$. As illustrated in Fig.~\ref{fig:db_vs_rate}, two scenarios can occur:
\begin{enumerate}[i)]
    \item if $\underline\lambda(L)< \hat{k}_{\sigma}$, then, at $\db=0$, the closed-loop poles lie on the real axis and the convergence rate is given by $\alpha_2$ in \eqref{eq:alpha2} with $\db=0$. As $\db$ increases, it always holds that $\underline\lambda(L)< \hat{k}_{\sigma} \leq k_{\sigma}$. Hence, the convergence rate continues to be $\alpha_2$. 
    \item if $\underline\lambda(L)\geq \hat{k}_{\sigma}$, then, at $\db=0$, the closed-loop poles lie on the asymptotes and the convergence rate is $-\hat\sigma=(d+\dt)/(2m)$. As $\db$ increases, the breakaway point shifts leftward. For moderate values of $\db$, the closed-loop poles remain on the asymptotes and the convergence rate increases linearly with $\db$ according to $\alpha_1$ in \eqref{eq:alpha1}. When $\db$ becomes large enough that $\underline\lambda(L)< k_{\sigma}$, the closed-loop poles move on the real axis and the convergence rate transitions to $\alpha_2$ in \eqref{eq:alpha2}.
\end{enumerate}

Combining Fig.~\ref{fig:db_vs_damping} and Fig.~\ref{fig:db_vs_rate}, we see that if $\underline\lambda(L)$ is sufficiently small so that the case in Fig.~\ref{fig:db_vs_rate}(a) occurs, a trade-off arises between the damping ratio and the convergence rate. In contrast, if $\underline\lambda(L)\geq \hat{k}_{\sigma}$, then increasing $\db$ in a reasonable range improves both quantities simultaneously. Of course, excessively large values of $\db$ are not desirable. 

For controller parameter tuning, one can first determine which case in Fig.~\ref{fig:db_vs_rate} applies to the given power system dynamics. Then, using the vertical axes of Fig.~\ref{fig:db_vs_damping} and Fig.~\ref{fig:db_vs_rate}, one can check whether the desired damping ratio and convergence rate are achievable. If they are, the value of $\db$ can be read from the horizontal axis.

\section{Numerical Illustrations}\label{sec:simulation}
In this section, we provide numerical illustrations for the  oscillation damping tuning process and control performance under FS control. 

\begin{figure}[htb]
\centering
\includegraphics[width=0.5\columnwidth]{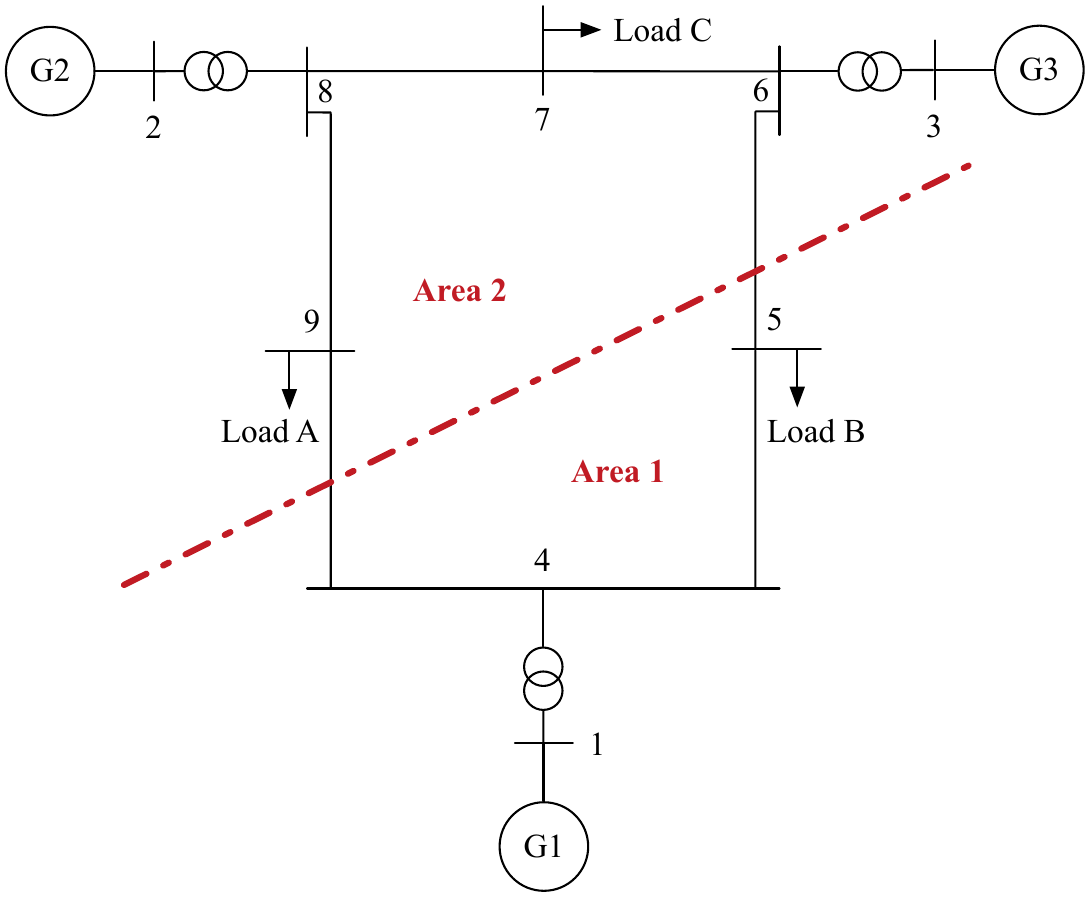}
\vspace{-5pt}
\caption{Diagram of $9$-bus $3$-generator WSCC test case.}
\label{fig:WCSS_test_case}
\end{figure}

\begin{figure}[t!]
\centering
\subfigure[System without inverters]
{\includegraphics[width=0.8\columnwidth]{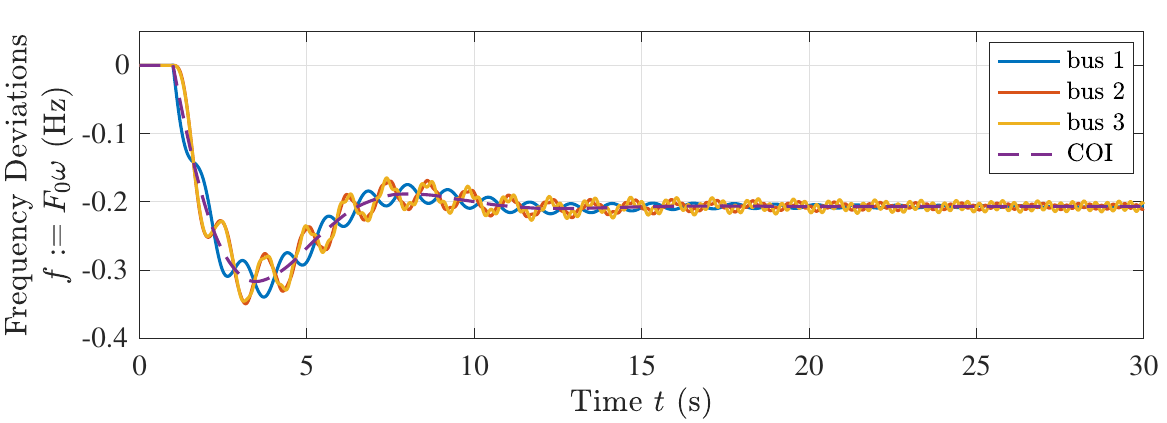}\label{fig:sw-simu}}
\hfil
\vspace{-5pt}
\subfigure[System under FS control with $d_{\mathrm{b}}=\SI{0}{\pu}$]
{\includegraphics[width=0.8\columnwidth]{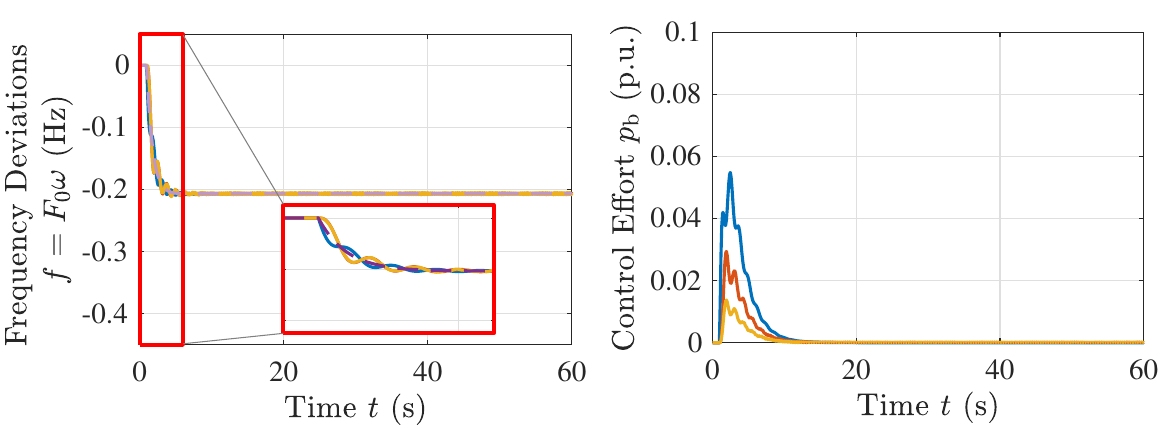}\label{fig:FS0}}
\hfil
\vspace{-5pt}
\subfigure[System under FS control with $d_{\mathrm{b}}=\SI{35.9}{\pu}$]
{\includegraphics[width=0.8\columnwidth]{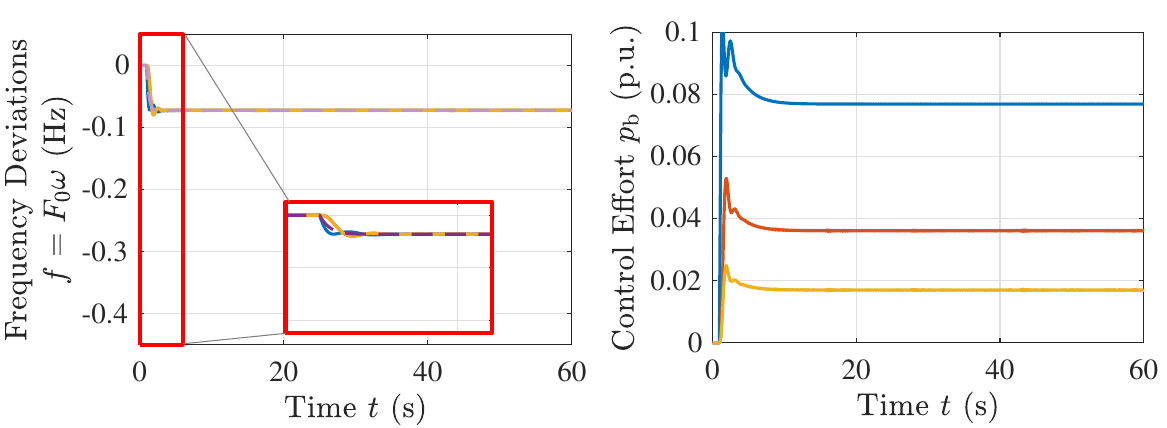}\label{fig:damp35}}
\vspace{-10pt}
\caption{Performance of WSCC system when a $\SI{-0.2}{\pu}$ step power change is introduced to bus $1$ at $t=\SI{1}{\second}$.}
\label{fig:compareS}
\end{figure}

The simulations are performed on the Western System Coordinating Council (WSCC) $9$-bus $3$-generator system in Fig.~\ref{fig:WCSS_test_case} taken from Power System Toolbox~\citep{chow1992toolbox}. In order to mimic the weakly connected network, we modify the existing WSCC test case by increasing the impedance of the lines $4-9$ and $5-6$ to $20$ times its value, which results in a distinct $2$-area system. The dynamic model is then built upon the
Kron reduced system where only the 3 generator buses
are retained. Even though our previous analysis is based on the proportionality assumption, the simulations are still conducted with the heterogeneous generator inertia and damping coefficients directly obtained from the dataset, i.e., $m_1 \!=\! \SI{27.28}{\second}$, $m_2 \!=\! \SI{12.80}{\second}$, $m_3 \!=\! \SI{6.02}{\second}$, $d_1 \!=\! \SI{9.6}{\pu}$, $d_2 \!=\! \SI{2.5}{\pu}$, $d_3 \!=\! \SI{1}{\pu}$. Given that the turbine coefficients are not provided by the dataset, we set the turbine inverse droops and time constants to $d_{\mathrm{t},1}=d_{\mathrm{t},2}=d_{\mathrm{t},3}=\SI{15}{\pu}$, and $\tau_1\!=\!\tau_2\!=\!\tau_3 \!=\! \SI{2}{\second}$, respectively. For the purpose of comparison, frequency deviations of this system without additional inverter control following a sudden $\SI{-0.2}{\pu}$ step change
in power consumption on bus $1$
at $t\!=\!\SI{1}{\second}$ is provided in Fig.~\ref{fig:sw-simu}, which exhibits a frequency drop.

We then add an inverter with FS control to each bus $i$, whose tuning relies on the knowledge
of a representative generator in some form although the proportionality
assumption is violated. We define the representative generator inertia constant
as the mean of individual generator inertia constants, i.e., $m := (\sum_{i=1}^3 m_i)/3=\SI{15.37}{\second}$. Accordingly, the proportionality parameters are given by $r_i \!:= m_i/m$. Then, we define the representative damping coefficient as $d \!:=\! (\sum_{i=1}^3 d_i)/(\sum_{i=1}^3 r_i)\!=\! \SI{4.37}{\pu}$ and the representative turbine inverse droop and time constant as $d_{\mathrm{t}} \!:=\! (\sum_{i=1}^3 d_{\mathrm{t},i})/(\sum_{i=1}^3 r_i)\!=\!\SI{15}{\pu}$ and $\tau \!:=\! (\sum_{i=1}^{3} \tau_i)/3\!=\!\SI{2}{\second}$, respectively. Considering that the steady-state frequency deviations of the original system without inverters are already satisfactory \citep{UCTLbook,NG2016standard}, we initially set $d_\mathrm{b}=\SI{0}{\pu}$. The performance of the system in this case is shown in Fig.~\ref{fig:FS0}, which successfully removes the frequency Nadir as expected. However, pronounced inter-area oscillations still exist, which motivates us to tune $d_\mathrm{b}$ further.

\begin{figure}[t!]
\centering
\includegraphics[width=0.8\columnwidth]{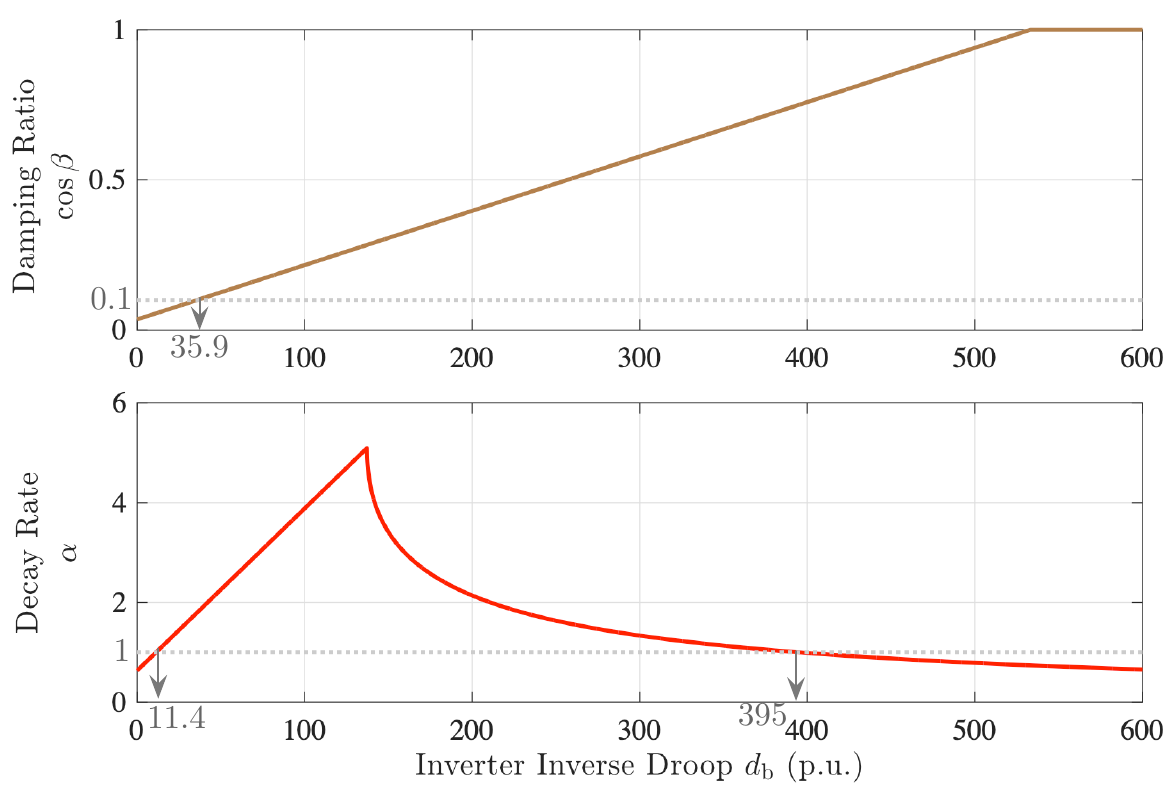}
\vspace{-10pt}
\caption{Damping ratio and decay rate as a function of $d_\mathrm{b}$.}
\label{fig:metric}
\end{figure}

To this end, we first plot the damping ratio and convergence rate as a function of $d_\mathrm{b}$ in Fig.~\ref{fig:metric}. Then, if we set the desired damping ratio $\cos\beta=0.1$ and convergence rate $\alpha=1$ to satisfy the grid code~\citep{WP2021}, then $d_\mathrm{b}\geq\SI{35.9}{\pu}$ and $\SI{11.4}{\pu}\leq d_\mathrm{b}\leq\SI{395.0}{\pu}$ are required, respectively, according to Fig.~\ref{fig:metric}. Thus, we need $\SI{35.9}{\pu}\leq d_\mathrm{b}\leq\SI{395.0}{\pu}$. Considering that a larger $d_\mathrm{b}$ usually implies a higher control effort, we finally choose $d_\mathrm{b}=\SI{35.9}{\pu}$ for oscillation damping. The performance of the system in this case is shown in Fig.~\ref{fig:damp35}, where the oscillations are damped to a satisfactory extent at the cost of higher control effort.

\section{Conclusion}\label{conclusion}
This paper has developed a root-locus-based method for analyzing the transient performance of inverter-based power systems under a proportional generator-inverter model. Sufficient conditions that guarantee the prescribed damping ratio and convergence rate under FS have been derived. The conditions indicate a structural property: the damping ratio depends only on the largest eigenvalue of a scaled network Laplacian, whereas the convergence rate is determined only by its smallest nonzero eigenvalue. Based on these conditions, we have provided guidelines for tuning the inverter controller parameter. Numerical simulations have been provided to demonstrate the effectiveness of the proposed approach.
Future work includes studying the transient performance for heterogeneous agents over directed networks.

\bibliography{bib}

\end{document}